\documentclass[12pt,a4paper]{article}

\usepackage{amsmath}
\usepackage{subfig}
\usepackage{graphicx}
\usepackage{amssymb}
\usepackage{epstopdf}
\usepackage{appendix}
\usepackage{enumerate}
\usepackage{color}
\usepackage{authblk}

\usepackage{float}
\usepackage{subfloat}
\newcommand{\nc}{\newcommand}
\nc{\eq}{\begin{equation}}
\nc{\eeq}{\end{equation}}
\nc{\eqa}{\begin{eqnarray}}
\nc{\eeqa}{\end{eqnarray}}
\nc{\ar}{\begin{array}}
\nc{\ear}{\end{array}}
\nc{\bfig}{\begin{figure}}
\nc{\efig}{\end{figure}}
\nc{\dg}{\dagger}
\nc{\eps}{\frac{\epsilon}{2}}
\nc{\juuri}{\sqrt{\Omega^2+(\eps)^2}}
\nc{\sx}{\sigma_x}
\nc{\sy}{\sigma_y}
\nc{\sz}{\sigma_z}
\nc{\spl}{\sigma_+}
\nc{\sm}{\sigma_-}
\nc{\Sx}{\bar{\sigma}_x}
\nc{\Sy}{\bar{\sigma}_y}
\nc{\Sz}{\bar{\sigma}_z}
\nc{\Spl}{\bar{\sigma}_+}
\nc{\Sm}{\bar{\sigma}_-}
\nc{\nn}{\nonumber}
\nc{\noi}{\noindent}
\nc{\omt}{\tilde{\omega}}
\nc{\Somt}{S(\omt)}
\nc{\Somtd}{S^{\dg}(\omt)}
\nc{\got}{\gamma_{\omega}(t)}
\nc{\gmot}{\gamma_{-\omega}(t)}
\nc{\po}{\mathcal{P}}
\nc{\qo}{\mathcal{Q}}
\nc{\adg}{a^{\dg}}
\nc{\gammat}{\tilde{\gamma}}
\nc{\Q}{$\mathcal{Q}$}
\nc{\C}{$\mathcal{C}$}
\nc{\kvec}{\mathbf{k}}

\def\bra#1{\mathinner{\langle{#1}|}}
\def\ket#1{\mathinner{|{#1}\rangle}}

\newcommand{\qed}{\nobreak \ifvmode \relax \else
      \ifdim\lastskip<1.5em \hskip-\lastskip
      \hskip1.5em plus0em minus0.5em \fi \nobreak
      \vrule height0.75em width0.5em depth0.25em\fi}

\title{\vspace{-1.0cm} \textbf{Frozen and Invariant Quantum Discord under Local Dephasing Noise}}
\date{}
\author[1]{G\"{o}ktu\u{g} Karpat}
\affil[1]{\small Faculdade de Ci\^encias, UNESP - Universidade Estadual Paulista, Bauru, SP, 17033-360, Brazil}
\author[2]{Carole Addis}
\affil[2]{\small SUPA, EPS/Physics, Heriot-Watt University, Edinburgh, EH14 4AS, UK}
\author[3,4]{Sabrina Maniscalco}
\affil[3]{\small Turku Center for Quantum Physics, Department of Physics and Astronomy, University of Turku, FIN-20014, Turun yliopisto, Finland}
\affil[4]{\small Centre for Quantum Engineering, Department of Applied Physics,  School of Science, Aalto University, P.O. Box 11000, FIN-00076 Aalto, Finland}

\begin{document}
\maketitle

\section{Introduction}

In nature, there exist various different types of correlations among the constituents of composite physical systems. While macroscopic systems only form correlations of classical nature, quantum mechanics allows for the existence of curious correlations devoid of a classical analogue, such as quantum entanglement. The idea of entanglement is as old as the quantum theory itself. In fact, Schr\"{o}dinger himself believed that entanglement is not one but rather the characteristic trait of quantum mechanics. Besides its foundational importance for quantum mechanics, entanglement has attracted renewed interest in the quantum physics community during the past few decades. The reason lies in the fact that the concept of entanglement has emerged as the main resource of quantum information science \cite{ent1}-\cite{ent3}, that is, it has been shown to be fundamental to the applications of quantum information processing, quantum computation and quantum cryptography.

Despite the central role of entanglement in quantum information science, recent investigations have proved that it might not be the only kind of quantum correlation serving as a resource for quantum information tasks. For instance, it has been demonstrated that some quantum systems in separable states can perform more efficiently than their classical counterparts in certain applications \cite{sep}. As a consequence, numerous different correlation measures have been introduced in the recent literature to be able to characterize and quantify the non-classical correlations in quantum systems \cite{discord1}-\cite{discord6}, which cannot be captured by entanglement measures. Among them, quantum discord, originally proposed by Ollivier and Zurek \cite{discord1} and independently by Henderson and Vedral \cite{discord2} as a measure of the quantumness of correlations, has received remarkable attention. This is mainly due to its usefulness as a resource in various quantum protocols such as the distribution of entanglement \cite{discorduse1}-\cite{discorduse1c}, quantum locking \cite{discorduse2}, entanglement irreversibility \cite{discorduse3} and many others \cite{discorduse4}-\cite{discorduse4b}.

Realistic quantum mechanical systems unavoidably interact with their surrounding environments. The effects of this inevitable interaction between the principal system of interest and its environment can be understood within the framework of open quantum systems theory \cite{bp}-\cite{qtoc}. Indeed, such interactions are typically detrimental for the crucial quantum traits present in the principal system as environment induced decoherence quickly destroys them in general. Therefore, protecting the precious quantum resources against the effects of the environment is of great importance, and also a major challenge for the realization of quantum computing devices.

One of the most striking outcomes of the decoherence process is the total loss of entanglement in composite systems in a finite time interval, also known as the sudden death of entanglement \cite{Sudden, Sudden2}. On the other hand, quantum discord has been shown to be robust against sudden death \cite{dissud1}-\cite{dissud4}, that is, where entanglement vanishes in finite time, quantum discord disappears only asymptotically. Another remarkable result related to quantum discord is the existence of a sharp transition between the loss of classical and quantum correlations during the time evolution of the open system \cite{m1}-\cite{m2}. In particular, this phenomenon implies that, under a suitable local decoherence setting, quantum correlations as quantified by quantum discord remains frozen at its initial value for a certain time interval, unaffected by the destructive effects of the environment, before they finally start to decay. Moreover, considering non-Markovian open system models, where the role of memory effects in the dynamics is no longer neglected, several intervals of frozen quantum discord throughout the dynamics have been observed \cite{m3}. In fact, assuming a particular sort of local non-Markovian environment model with coherence trapping, it has also been demonstrated that quantum discord can be forever frozen during the whole time evolution thus becoming time invariant \cite{nm1}. Although it has also been recently shown that entanglement might also become time invariant under suitable conditions, this requires the existence of global environmental interactions and decoherence free subspaces \cite{frozenent1}-\cite{frozenent2}.

In this chapter, we intend to explore and review some remarkable dynamical properties of quantum discord under various different open quantum system models. Specifically, our discussion will include several concepts connected to the phenomena of time invariant and frozen quantum discord. Furthermore, we will elaborate on how these two phenomena are related to both  the non-Markovian features of the open system dynamics and the usage of dynamical decoupling protocols.

\section{Quantum Discord}

Correlations of purely quantum nature and more general than entanglement can be quantified by quantum discord \cite{discordreview}, which is defined as the discrepancy between the two natural yet non-identical quantum generalizations of classically equivalent expressions for mutual information. Assuming that one has two classical random variables denoted by $A$ and $B$, the first classical expression for the mutual information is given as follows
\begin{align}
I(A:B)=H(A)+H(B)-H(A,B),
\end{align}
where the Shannon entropy $H(X)=-\sum_x p_x \log_2 p_x$ measures the amount of information one can obtain on average after learning the outcome of the measurement of the random variable $X\in(A,B)$, and the probability of obtaining the outcome labelled by $x$ is denoted as $p_x$. Here, $H(A,B)=-\sum_{x,y} p_{x,y} \log_2 p_{x,y}$ is the joint entropy of the random variables $A$ and $B$, which quantifies the total amount of ignorance about the pair, and $p_{x,y}$ is their joint probability distribution. On the other hand, the second definition of the mutual information can be given as
\begin{align}
J(A:B)=H(A)-H(A|B),
\end{align}
where $H(A|B)$ is the entropy of the random variable $A$ conditioned on the outcome of $B$. Considering that $H(A|B)=H(A,B)-H(B)$ in classical information theory, then it is not difficult to observe that both definitions of the classical mutual information are equivalent.

A rather straightforward quantum analogue of the classical mutual information, the so-called quantum mutual information, can be deduced by replacing the probability distributions related to the random variables with density operators. In particular, while the reduced density matrices $\rho_A=\text{Tr}_B\{\rho_{AB}\}$ and $\rho_B=\text{Tr}_A\{\rho_{AB}\}$ take the place of the probability distributions for the random variables $A$ and $B$ respectively, $\rho_{AB}$ replaces the joint probability distribution for the $A,B$ pair. Moreover, the Shannon entropy is replaced by the von Neumann entropy $S(\rho)=-\text{Tr} (\rho\log\rho)$. With these considerations, one can define the quantum mutual information in the following way
\begin{align}
\mathcal{I}(\rho_{AB})=S(\rho_A)+S(\rho_B)-S(\rho_{AB}),
\end{align}
which, although first named by Cerf and Adami \cite{cerf}, is known to have already been considered by Stratonovich \cite{Stratonovich} many years ago. Today, it is widely agreed in the literature that the quantum mutual information is the information-theoretic quantifier of total correlations in a bipartite quantum state. This is also supported by several operational interpretations. For instance, Groisman \emph{et al} have argued that quantum mutual information can be defined via the amount of noise that is required to completely destroy the correlations present in the system  \cite{Groisman}. Another supportive example has been provided by Schumacher and Westmoreland, where they have shown that if two separated parties share a correlated quantum system to be used as the key for a ``one-time pad cryptographic system", the maximum amount of information that one party can send securely to the other is equal to the quantum mutual information of the shared correlated state \cite{SW}.
  
The other quantum analogue of the classical mutual information is based on the concept of conditional entropy. By definition, classical conditional entropy $H(A|B)$ depends on the outcome of the random variable $B$. However, as is well known in quantum theory, measurements  generally do disturb quantum systems. In other words, performing measurements on the system $B$ affects our knowledge of the system $A$, and how much the system $A$ is disturbed by a measurement of $B$ depends on the type of measurement performed, i.e., on the measurement basis.
Here, we assume that the measurements on the system $B$ is of von Neumann type (projective measurements), i.e., they can be described by a complete set of orthonormal projectors $\{\Pi_{k}^{B}\}$ corresponding to the outcome $k$. Then, the quantum generalization of the classical mutual information in terms of the conditional entropy now reads
\begin{align}
\mathcal{J}(\rho_{AB})=S\left(\rho_{A}\right)-\sum_{k}p_{k}S\left(\rho_{A|k}, \right),
\end{align}
where $\rho_{A|k}=Tr_B(\Pi_{k}^{B}\rho_{AB}\Pi_{k}^{B})/p_k$ is the remaining state of the subsystem $A$ after obtaining the outcome $k$ with probability $p_k=Tr_{AB}(\Pi_{k}^{B}\rho_{AB}\Pi_{k}^{B})$ in the subsystem $B$. Supposing that the reduced state of the subsystem to be measured here is a two-level quantum system, one can construct the local von Neumann measurement operators $\{\Pi_{1}^{B}, \Pi_{2}^{B}\}$ in the following way:
\begin{align}
\Pi_{1}^{B} =\frac{1}{2}\left(I^{B}_{2}+\sum_{j=1}^{3}n_{j}\sigma_{j}^{B}\right),
\hspace{0.8cm} \Pi_{2}^{B} =\frac{1}{2}\left(I^{B}_{2}-\sum_{j=1}^{3}n_{j}\sigma_{j}^{B}\right),
\end{align}
where $\sigma_{j}(j=1,2,3)$ are the Pauli spin operators and $n=(\sin\theta\cos\phi,\sin\theta\sin\phi,\cos\theta)^{T}$ is a unit vector on the Bloch sphere with $\theta \in [0,\pi)$ and $\phi \in [0,2\pi)$. It should be quite clear now that the two classically equivalent definitions of the mutual information do not agree in the quantum realm in general, as $\mathcal{J}(\rho_{AB})$ is basis dependent and reduces to $\mathcal{I}(\rho_{AB})$ only under special conditions.

A reasonable assumption is that the total amount of correlations in a bipartite quantum system can be divided into two parts and written as the sum of quantum ($\mathcal{Q}$) and classical ($\mathcal{C}$) correlations,
\begin{align}
\mathcal{I}(\rho_{AB})=\mathcal{Q}(\rho_{AB})+\mathcal{C}(\rho_{AB}).
\end{align}
It has been recently suggested that the classical correlations in a bipartite quantum system can be quantified by the maximization of $\mathcal{J}(\rho_{AB})$ over all projective measurements on the subsystem $B$, that is, $\mathcal{C}(\rho_{AB})=\max_{\Pi_{k}^{B}}\{ \mathcal{J}(\rho_{AB})\}$ \cite{discord2}. This quantity captures the maximum amount of information that can be obtained about the subsystem $A$ by performing local measurements on the subsystem $B$, i.e., the locally accessible information. Consequently, genuine quantum correlations can be measured by subtracting the classical part of the correlations from the total amount, 
\begin{align} \label{discord}
\mathcal{Q}(\rho_{AB})&=\mathcal{I}(\rho_{AB})-\max_{\Pi_{k}^{B}}\{ \mathcal{J}(\rho_{AB})\}, \nonumber \\
&=\mathcal{I}(\rho_{AB})-\max_{\Pi_{k}^{B}} \left\lbrace  S\left(\rho_{A}\right)-\sum_{k}p_{k}S\left(\rho_{A|k}, \right) \right\rbrace,
\end{align}
which is the well known quantum discord \cite{discord1}, also known as the locally inaccessible information. Quantum discord quantifies the genuine quantum correlations in a bipartite quantum system. It is important to emphasize that quantum discord is more general than entanglement in the sense that it is possible for some separable mixed states to have non-zero quantum discord and thus non-classical correlations. For instance, despite the fact that the state $\rho=1/2(\ket{0}\bra{0}_A\otimes\ket{-}\bra{-}_B+\ket{+}\bra{+}_A\otimes\ket{1}\bra{1}_B)$ with $\ket{\pm}=(\ket{0}\pm\ket{-})/\sqrt{2}$ has zero entanglement, it still cannot be described by classical means, i.e., by a classical bivariate probability distribution. The reason lies in the non-orthogonality of the reduced states of the subsystems $A$ and $B$, which guarantees the impossibility of locally distinguishing the states of each subsystem. Indeed, in order to evaluate classical correlations $\mathcal{J}(\rho_{AB})$ and thus quantum discord $ \mathcal{Q}(\rho_{AB})$, one might more generally perform the optimization over all possible positive operator valued measures (POVMs) instead of the set of projective measurements. Nevertheless, projective measurements are most widely considered in the literature since, even for this simpler case, there exists no available generic analytical expression for quantum discord and analytical results have been obtained only in few restricted cases, such as Bell-diagonal \cite{luo} or X shaped states of two qubits \cite{ali}. On the other hand, it has also been shown that projective measurements are almost sufficient for calculating the quantum discord of two qubits, and they are always optimal for the case of rank-2 states \cite{almostpro}.

Let us lastly mention some of the important properties of quantum discord. Being a measure for correlations, it is expectedly non-negative, which is due to the concavity of the conditional entropy \cite{entcov}. One can also notice that quantum discord is not a symmetric quantity in general, meaning that its value depends on whether the measurement is performed on subsystem A or subsystem B, which is a result of the asymmetry of the conditional entropy. It is invariant under local change of basis, that is, invariant under local unitary transformations. Furthermore, for pure states, quantum discord becomes a measure of entanglement being reduced to entanglement entropy, and it vanishes if and only if the considered state is classical-quantum \cite{discord1}. Finally, even though there are now numerous discord-like or related quantum correlation measures introduced in the literature \cite{discord6}, we will mainly focus on the original entropic discord in this chapter.

\section{Open System Dynamics of Quantum Discord}

In this section, we will explore some remarkable dynamical features of quantum discord for bipartite systems under local dephasing noise. Particularly, we will discuss the phenomena of frozen and time-invariant quantum correlations in case of both Markovian and non-Markovian interactions between the open quantum system and its environment. We will also review the role of dynamical decoupling protocols in protecting quantum correlations from environment induced decoherence.

\subsection{Frozen Quantum Discord}

\subsubsection{Decoherence of Classical and Quantum Correlations}

In the following, we will consider a bipartite system of two non-interacting qubits evolving in time under the action of a locally acting dephasing (or equivalently phase-flip) channel. One can start from a Markovian time-local master equation to obtain the dynamics of the bipartite open system. For each of the two-level systems, the Markovian master equation is given by
\begin{align}
\dot{\rho}_{A(B)}=\frac{\gamma}{2}[\sigma_z^{A(B)}\rho_{A(B)}\sigma_z^{A(B)}-\rho_{A(B)}],
\end{align}
where $\sigma_z^{A(B)}$ is the usual Pauli spin operator in the z-direction acting on the subsystem $A(B)$. Using the solution of the above master equation, the time evolution of the open system in such a decoherence scenario can be expressed with the help of the operator-sum representation as follows
\begin{align} \label{Kdynamics}
\rho_{AB}(t)= \sum_{i,j=1}^2 (M_i^{A} \otimes M_j^{B}) \rho_{AB}(0) (M_i^{A} \otimes M_j^{B})^\dagger,
\end{align}
where the Kraus operators $M_1$ and $M_2$ defining the dynamical dephasing map are given as
\begin{align}
M^{A(B)}_1(t)= \begin{pmatrix} \sqrt{1-p(t)/2} & 0\\ 0 & \sqrt{1-p(t)/2} \end{pmatrix}, \qquad
M^{A(B)}_2(t)= \begin{pmatrix} \sqrt{p(t)/2} & 0 \\ 0 & -\sqrt{p(t)/2} \end{pmatrix},
\end{align}
and the explicit time dependence of the dephasing factor is $p(t)=1-\exp(-\gamma t)$ with $\gamma$ being the dephasing rate. Note that here the local noise channels act on both qubits in the same way with identical dephasing factors. In Ref. \cite{m1}, the dynamics of quantum discord and classical correlations have been discussed for a system of two qubits under such a local dephasing setting.

For this purpose, one can consider the local dephasing dynamics of the Bell-diagonal states, i.e., the class of two-qubit states with maximally mixed reduced density operators,
\begin{align}
\rho_{AB}=\frac{1}{4}\left(\mathbb{I}_{AB}+\sum_{i=1}^3c_i\sigma_i^A\otimes\sigma_i^B\right),
\end{align}
where $I_{AB}$ is the $4\times4$ identity matrix and $c_j$ are real numbers such that $0\leq |c_j| \leq1$. This class of states are indeed nothing but the convex combination of the the four Bell states given as
\begin{align} \label{Bdiag}
\rho_{AB}= \lambda^+_\phi\ket{\phi^+}\bra{\phi^+} + \lambda^+_\psi\ket{\psi^+}\bra{\psi^+} + \lambda^-_\psi\ket{\psi^-}\bra{\psi^-} + \lambda^+_\phi\ket{\phi^+}\bra{\phi^+}
\end{align}
where the non-negative eigenvalues of the bipartite density matrix $\rho_{AB}$ read
\begin{align}
\lambda^{\pm}_\phi=[1\pm c_1\mp c_2+c_3]/4,  \hspace{1cm}
\lambda^{\pm}_\psi=[1\pm c_1\pm c_2-c_3]/4, \label{eig}
\end{align}
and $\ket{\phi^\pm}=(\ket{00}\pm\ket{11})/\sqrt{2}$, $\ket{\psi^\pm}=(\ket{01}\pm\ket{10})/\sqrt{2}$ are the four maximally entangled Bell states. From a geometrical perspective, the class of Bell-diagonal states can be thought to form a tetrahedron with the four maximally entangled Bell states sitting in the extreme points. In this case, dynamics of the mutual information $\mathcal{I}(\rho_{AB}(t))$, the classical correlations $\mathcal{C}(\rho_{AB}(t))$ and the quantum correlations measured by quantum discord $\mathcal{Q}(\rho_{AB}(t))$ are given by \cite{m1}
\begin{align}
\mathcal{I}(\rho_{AB}(t)) &= 2+\sum_{kl}\lambda_k^l(t)\log_2\lambda_k^l(t), \label{mutinf} \\
\mathcal{C}(\rho_{AB}(t)) &= \sum_{j=1}^2\frac{1+(-1)^j\mathcal{X}(t)}{2}\log_2[1+(-1)^j\mathcal{X}(t)]\label{clas} \\
\mathcal{Q}(\rho_{AB}(t)) &= 2+\sum_{kl}\lambda_k^l(t)\log_2\lambda_k^l(t) -  \sum_{j=1}^2\frac{1+(-1)^j\mathcal{X}(t)}{2}\log_2[1+(-1)^j\mathcal{X}(t)] \label{quant},
\end{align}
where $\mathcal{X}(t)=\max\{|c_1(t)|,|c_2(t)|,|c_3(t)|\}$, $k=\phi,\psi$, and $l=\pm$. We should stress that, since the Bell diagonal states preserve their form under the local dephasing map, time dependence of the eigenvalues $\lambda_k^l$, and thus the coefficients $c_i$, can be easily obtained from the time evolved density matrix $\rho_{AB}(t)$ as $c_1(t)=c_1(0)\text{exp}(-2\gamma t)$, $c_2(t)=c_2(0)\text{exp}(-2\gamma t)$, and $c_3(t)=c_3(0)=c_3$.

\begin{figure}
\centering
\includegraphics[width=0.5\textwidth]{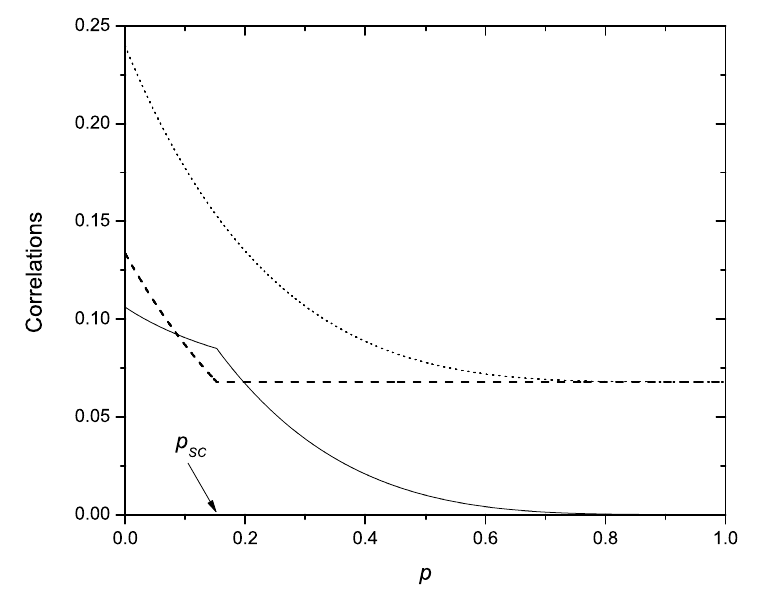} 
\caption{Classical correlations $\mathcal{C}(\rho_{AB}(t))$ (dashed line), quantum correlations $\mathcal{Q}(\rho_{AB}(t))$ (solid line), and total correlations $\mathcal{I}(\rho_{AB}(t))$ (dotted line) versus the parametrized time $p(t)$ under local dephasing noise. The chosen initial Bell diagonal state parameters are $c_1 = 0.06, c_2 = 0.42$, and $c_3 = 0.30$, and the sudden change occurs at the point $p_{sc} = 0.15$. Note that the quantum discord $\mathcal{Q}(\rho_{AB}(t))$ is greater than the classical correlations $\mathcal{C}(\rho_{AB}(t))$ for $0.09 \leq p \leq 0.20$.}
\label{fig1}
\end{figure}

The authors of Ref. \cite{m1} have identified three fundamentally different types of dynamical behavior for the correlations depending on the region of parameters $c_i$ of the initial Bell-diagonal state. In the first region where $|c_3|>|c_1|,|c_2|$, classical correlations $\mathcal{C}(\rho_{AB}(t))$ stay constant independently of the dephasing parameter $p(t)$ and quantum discord $\mathcal{Q}(\rho_{AB}(t))$ decays in a monotonic fashion. In the second region in which $|c_3|=0$, we have a monotonic decay for both classical and quantum correlations, which is not particularly interesting. Lastly, and most interestingly, in the region of parameters where $|c_1|>|c_2|,|c_3|$ or $|c_2|>|c_1|,|c_3|$, after decaying monotonically until a specific parametrized time $p_{SC}$, classical correlations $\mathcal{C}(\rho_{AB}(t))$ experiences a sudden change in their dynamics and remain constant afterwards. At the same specific time $p_{sc}$, quantum discord exhibits an abrupt change in its decay trend, and then continues to decay monotonically over time. An example of this scenario is displayed in Fig. \ref{fig1}. We note that such a sudden change in the dynamics had not been observed before in the literature for the other known quantum correlation measures. In addition, another interesting point we can observe here is that the early conjecture that $\mathcal{C}\geq \mathcal{Q}$ for any quantum state \cite{conject} is shown to be invalid, as can be clearly seen in Fig. \ref{fig1}.

Finally, the result that the classical correlations might become unaffected from the detrimental effects of the environment induced decoherence for a certain noise channel can be in principle used to simplify the evaluation of classical correlations and thus quantum discord, by removing the typically difficult optimization procedure in their definitions. More specifically, if one knows the dynamical map (assuming that such a map actually exists), for a given initial bipartite quantum state which preserves the classical correlations in the system, quantum discord will be given by the difference between the mutual information $\mathcal{I}(\rho_{AB})$ and the completely decohered mutual information $\mathcal{I}(\rho_{AB}|_{p=1})$, that is,
\begin{align}
\mathcal{Q}(\rho_{AB})=\mathcal{I}(\rho_{AB})-\mathcal{I}(\rho_{AB}|_{p=1}),
\end{align}
due to the simple fact that $\mathcal{I}(\rho_{AB})=\mathcal{Q}(\rho_{AB})+\mathcal{C}(\rho_{AB})$ and also $C(\rho_{AB})=\mathcal{I}(\rho_{AB}|_{p=1})$ in this case.

\subsubsection{Sudden transition between classical and quantum decoherence}

Following the discussion of frozen and time-invariant behavior of classical correlations under non-dissipative dephasing noise, we will now turn our attention to the possibility of observing frozen quantum correlations as measured by quantum discord. While it was first thought that quantum discord decays and vanishes asymptotically under Markovian noise, the authors of Ref. \cite{m2} remarkably identified a class of two-qubit initial states for which quantum discord preserves its initial value and therefore becomes frozen for a finite time interval until a critical time point is reached. This striking result, valid for dephasing noise, was the first evidence that  precious quantum features in open quantum systems can remain intact naturally even in presence of decoherence. Clearly, the existence of the phenomenon of frozen quantum discord might be highly important for quantum information protocols that rely on it as a resource.

\begin{figure}
\centering
\includegraphics[width=0.55\textwidth]{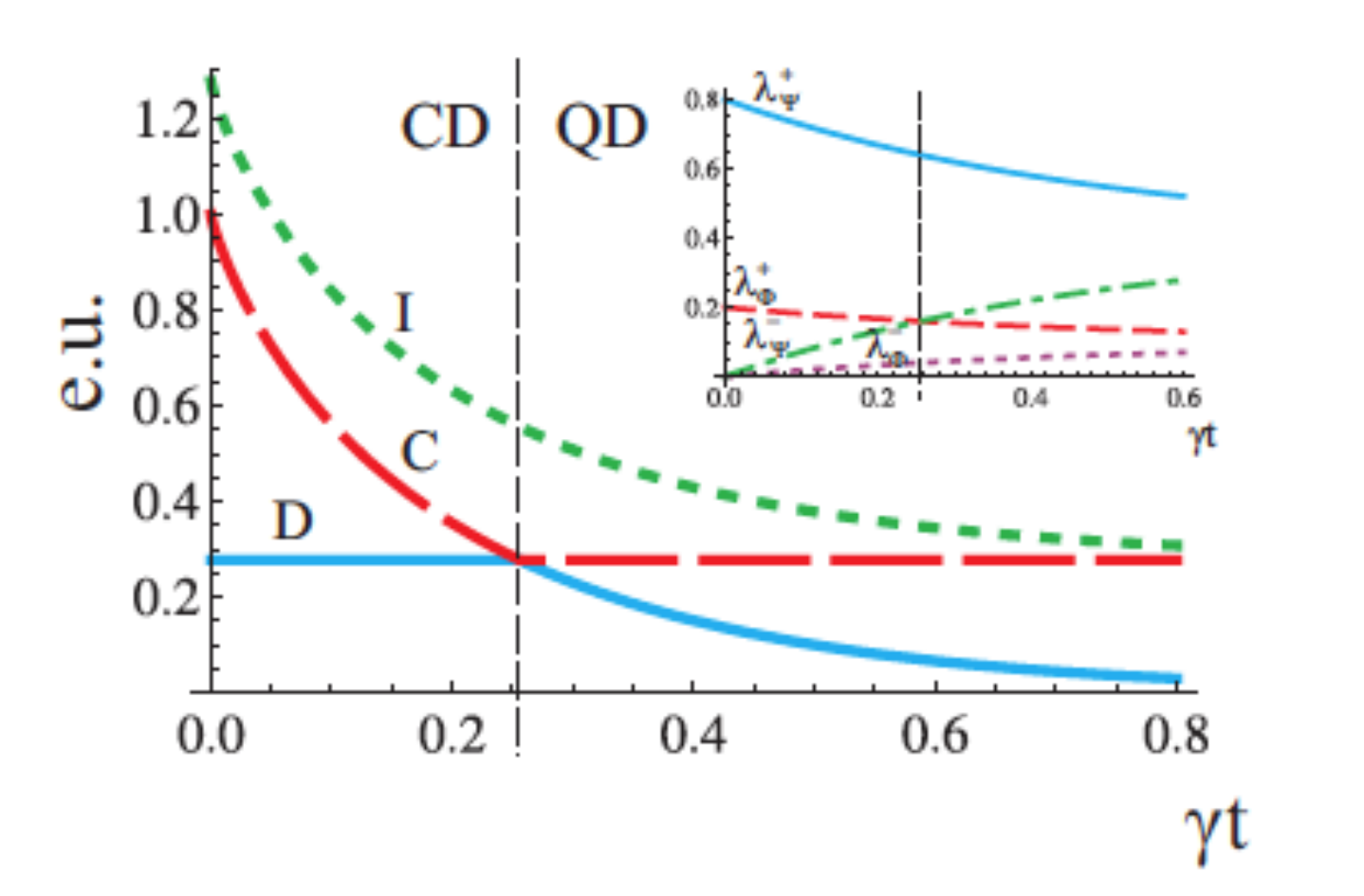} 
\caption{Dynamics of mutual information (green dotted line), classical correlations (red dashed line) and quantum discord (blue solid line) as a function of $t$ for $c_1(0)=1$, $c_2(0)=-c_3$ and $c_3=0.6$.  In the inset, we plot the eigenvalues, $\lambda^+_\psi$ (blue solid line), $\lambda^-_\psi$ (green dash dotted line), $\lambda_\phi^+$ (red dashed line) and $\lambda_\phi^-$ (violet dotted line) as a function of $\gamma t$ for the same parameters.}
\label{fig2}
\end{figure}

Similarly to the previous subsection, we consider a system of two non-interacting qubits in Bell-diagonal states under a locally acting Markovian dephasing noise. Consequently, the formulas for the classical correlations $\mathcal{C}(\rho_{AB}(t))$ and the quantum discord $\mathcal{Q}(\rho_{AB}(t))$ are once again given by Eq. (\ref{clas}) and Eq. (\ref{quant}), respectively. However, one can alternatively focus on a special class of initial states with parameters $c_1(0)=\pm 1$, $c_2(0)=\mp c_3(0)$ and $|c_3|<1$, which can be written as
\begin{align}
\rho_{AB}=\frac{(1+c_3)}{2}\ket{\Phi^{\pm}}\bra{\Phi^{\pm}}+\frac{(1-c_3)}{2}\ket{\Psi^{\pm}}\bra{\Psi^{\pm}}
\end{align}
For this class of initial Bell-diagonal states, it is straightforward to see from Eq. (\ref{mutinf}) that the quantum mutual information $\mathcal{I}[\rho_{AB}(t)]$ takes the following form:
\begin{align}
\mathcal{I}(\rho_{AB}(t))=\sum_{j=1}^2\frac{1+(-1)^jc_3}{2}\log_2[1+(-1)^jc_3]+\sum_{j=1}^2\frac{1+(-1)^jc_1(t)}{2}\log_2[1+(-1)^jc_1(t)]. \label{CI}
\end{align}
Recalling that $c_1(t)=\exp(-2\gamma t)$ and the form of the classical correlations $\mathcal{C}(\rho_{AB}(t))$ given in Eq. (\ref{clas}), it follows that, in case one has $t<\tilde t=-\text{ln}(|c_3|)/(2\gamma)$, the second term on the right-hand side of Eq. (\ref{CI}) is equal to the classical correlations $\mathcal{C}(\rho_{AB}(t))$ due to the fact that $|c_1(t)|>|c_2(t)|$ and $|c_3(t)|=|c_3|$. Hence, the quantum discord is given by the first term on the right-hand side, which is independent of time. In other words, quantum correlations $\mathcal{Q}(\rho_{AB}(t))$ as quantified by quantum discord, become frozen until a critical time $\tilde t$ is reached. Furthermore, one can control the time interval in which quantum discord is constant in time by adjusting the value of $|c_3|$. However, a trade-off exists, that is, the longer the time interval of frozen quantum correlations, the smaller the initial value of quantum discord one can obtain. Fig. \ref{fig2} shows the dynamical behavior of the quantum discord, the classical correlations and the mutual information for the initial Bell-diagonal state with parameters $c_1(0)=1$, $c_2(0)=-c_3$ and $c_3=0.6$. Here, one can observe that until the critical time $\tilde t$ only the classical correlations decay and quantum discord remains frozen. After this critical instant, where a sudden change occurs, classical correlations freeze and quantum discord begins to decay asymptotically. This is recognized as the phenomenon of sudden transition between classical and quantum decoherence \cite{m2}.

One can further investigate the roots of this curious sudden transition between classical and quantum decoherence from a geometrical point of view. To this aim, it is customary to adopt a quantity to measure the distance between quantum states. Here, following the approach of Ref. \cite{discord3}, we choose the relative entropy defined by $S(\rho_1||\rho_2)=-\text{Tr}(\rho_1  \log \rho_2)-S(\rho_1)$ as an entropic distance measure, although it cannot be technically considered as a distance. An alternative definition of the quantum discord can then be given as the distance of the considered state from the closest classical state $\rho_\text{cl}$. It turns out that the closest classical state to our system of choice given in Eq. (\ref{Bdiag}), i.e. Bell-diagonal states, during the open system dynamics will be given by \cite{discord3}
\begin{align} \label{closest}
\rho_\text{cl}(t)=\frac{q(t)}{2}\sum_{i=1,2}\ket{\Psi_i}\bra{\Psi_i}+\frac{1-q(t)}{2}\sum_{i=3,4}\ket{\Psi_i}\bra{\Psi_i}
\end{align}
where $q(t)=\lambda_1(t)+\lambda_2(t)$, and $\lambda_1(t)$ and $\lambda_2(t)$ are the two greatest time dependent eigenvalues of $\rho_{AB}$ given by Eq. (\ref{eig}) with $\ket{\Psi_i}$ being the corresponding Bell states. The inset of Fig. \ref{fig2} shows the time evolution of the eigenvalues $\lambda^\pm_\psi$ and $\lambda^\pm_\phi$, which can be considered as weights of the four Bell states forming the Bell-diagonal state. One can see that there is a transition time $\tilde t$ at which  the weight of $\ket{\phi_+}$ becomes identical to the weight of $\ket{\psi_-}$. This results in a sudden change in the second greatest eigenvalue. Thus, while the closest classical state for $t<\tilde t$ is given by
\begin{align}
\rho_\text{cl}(t<\tilde t)=\frac{1+e^{-2\gamma t}}{4}(\ket{\Psi^+}\bra{\Psi^+}+\ket{\Phi^+}\bra{\Phi^+}+\frac{1-e^{-2\gamma t}}{4}(\ket{\Phi^+}\bra{\Phi^+}+\ket{\Psi^+}\bra{\Psi^+},
\label{c1}
\end{align}
in case of $t>\tilde t$, the closest classical state will be given by
\begin{align}
\rho_\text{cl}(t>\tilde t)=\frac{1+c_3}{4}(\ket{\Psi^+}\bra{\Psi^+}+\ket{\Psi^-}\bra{\Psi^-}+\frac{1-c_3}{4}(\ket{\Phi^+}\bra{\Phi^+}+\ket{\Phi^-}\bra{\Phi^-}.
\label{c2}
\end{align}
\begin{figure}
\centering
\includegraphics[width=0.6\textwidth]{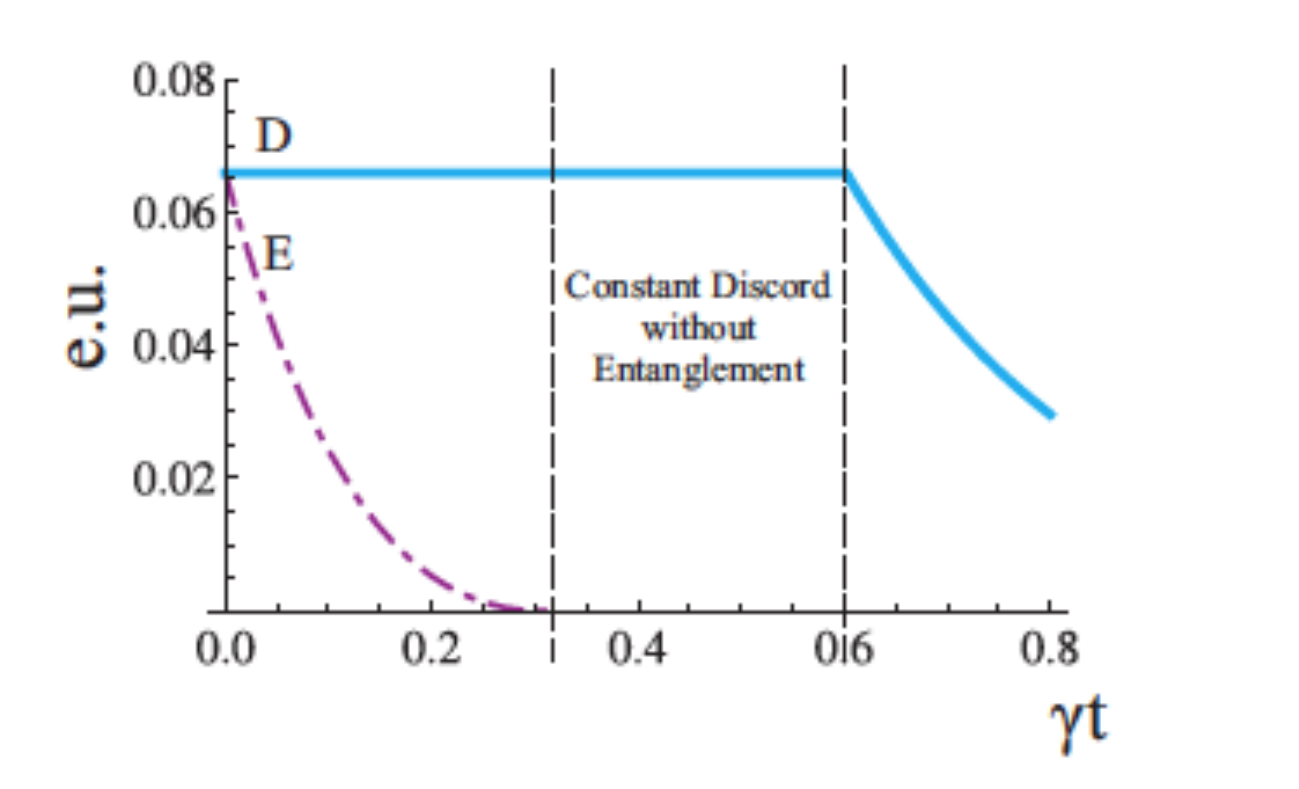} 
\caption{Dynamics of the entanglement (violet dashed-dotted line) and the quantum discord (blue solid line) as a function of $\gamma t$ for the initial state $c_1(0)=1$, $c_2(0)=-c_3$ and $c_3=0.6.$.}
\label{fig3}
\end{figure} 
Note that,
\begin{align}
D(\rho_{AB}(t)||\rho_\text{cl}(t))=-\text{Tr}(\rho_{AB}(t)\log_2\rho_\text{cl}(t))+\text{Tr}(\rho_{AB}(t)\log_2\rho_{AB}(t)),
\label{re}
\end{align}
from which it is straightforward to show that $D(\rho_{AB}||\rho_\text{cl})=\mathcal{D}(\rho_{AB})$, proving the equivalence of the original and relative entropy based definitions of quantum discord,  for the whole class of Bell-diagonal states. Based on this result, we see that the closest classical state to the considered one before the critical time at $\tilde t$ remains unchanged even though the state evolves in time. On the other hand, after the critical time is reached, the closest state starts to change over time and thus quantum discord exhibits a monotonically decaying behavior.

Moreover, to compare the dynamics of quantum discord and entanglement, it is instructive to study the relative entropy of entanglement $E$, which is defined as the distance to the closest separable state $\rho_S$. In this case, entanglement for the Bell-diagonal states can be calculated as
\eq
E=1+\lambda_1\log_2\lambda_1+(1-\lambda_1)\log_2(1-\lambda_1),
\eeq
where $\lambda_1$ is the greatest of the eigenvalues given by Eq. (\ref{eig}). Thus, entanglement decays in a monotonic way and vanishes when $t\geq t_s=-\text{ln}[(1-|c_3|)/(1+|c_3|)]/2\gamma$. More importantly, for $t_s<\tilde t$, entanglement vanishes even before quantum discord begins to decay, which occurs for $0<|c_3|<\sqrt{2}-1$. Figure \ref{fig3} displays an example of this situation, where one has constant quantum discord without entanglement, for the initial state with $c_1(0)=1$, $c_2(0)=-c_3$ and $c_3=0.6.$

After its first discovery, the phenomenon of frozen quantum discord has been explored in the literature in greater detail. For instance, necessary and sufficient conditions giving rise to the freezing of quantum discord has been explored for the Bell-diagonal states under local dephasing noise \cite{cen} and for certain multipartite states under other non-dissipative local noise channels \cite{sen}. Note that although the transition from constant to decaying quantum discord has been shown to occur abruptly for specific two-qubit states, it has been suggested that this transition might be sudden only for a idealized zero-measure subset of states within the whole set of two qubit states \cite{pinto}. The investigation of the dynamics of quantum discord in open quantum systems has not been limited to case of two level systems. In particular, frozen quantum discord has been also observed in hybrid qubit-qutrit states under local dephasing noise \cite{gedik}. Furthermore, due to the difficulties in the analytical evaluation of the original entropic quantum discord, numerous different geometric versions have been introduced to measure the distance of the considered state to the closest classical state according to a chosen metric. It has also been demonstrated that sudden change and freezing behavior of quantum discord can also be observed for geometric discord under non-dissipative local noise \cite{geo}. It is even possible to have double sudden transitions for a particular version of geometric discord, i.e., the trace-norm discord \cite{one,paula}. Later on, the phenomenon of frozen correlations has been shown to be a common feature of all bona fide measures of quantum correlations \cite{adesso}. Lastly, it has been proven that all geometric quantifiers of quantum correlations, having the properties of invariance under transposition, convexity, and contractivity under quantum channels, might lead to the freezing phenomenon under suitable local non-dissipative noise \cite{univ}.

On the other hand, the phenomena of sudden change and freezing of quantum discord have not remained as a purely theoretical construct. There have been several experiments demonstrating these peculiar effects with different physical systems. The first of these investigations has been performed by Xu et al. with an all optical experimental setup, where they have simulated the effects of a one-sided phase damping channel on a particular Bell-diagonal state \cite{dexp1}. In their experiment, the open systems is represented by a pair of photons that are entangled in their polarization degrees of freedom and generated by a spontaneous parametric downconversion process. The dephasing effect is produced when photons pass through a quartz plate of adjustable thickness which causes their polarization degrees of freedom to couples to their frequency degrees of freedom, acting as the environment of the photon. They have observed a clear evidence of the occurrence of the phenomenon of frozen quantum discord and also experimentally proved that quantum correlations can be greater than classical
correlations disproving the earlier conjecture. Another important experiment on the open system dynamics of quantum discord has been performed by Auccaise et al. using a room temperature nuclear magnetic resonance setup, where the environment induced sudden change takes place during the relaxation of two nuclear spins to the Gibbs state \cite{dexp2}. In their work, they have presented an evidence of both the sudden change and freezing of discord.

\subsubsection{Frozen discord in non-Markovian dephasing channels}

In this section, we will extend the investigation of the dynamics of classical and quantum correlations to the case where the bipartite open system is under the effect of  non-Markovian dephasing local noise channels. We will elaborate on the non-trivial consequences of the memory effects, emerging due to the non-Markovian nature of the noise, for the phenomenon of frozen quantum discord. To this end, we will consider a well established pure dephasing model where white noise producing Markovian evolution is replaced by colored noise giving rise to non-Markovian time evolution \cite{daffer}.

Let us now introduce the open quantum system model which describes the independent dephasing of two non-interacting qubits under local identical dephasing noise channels. The time evolution of each of the qubits can be given by a master equation of the form
\begin{align}
\dot{\rho}=K\mathcal{L}\rho, \label{master}
\end{align}
where $\rho$ is the density matrix of open quantum system, $\mathcal{L}$ is a Lindblad superoperator describing the time evolution of the open quantum system, and $K$ is a time-dependent integral operator acting on the system as $K\phi=\int_0^t k(t-t')\phi(t')dt'$. The function $k(t-t')$ is a memory kernel that determines the type of memory in the dynamics. Such a master equation emerges if one considers the time evolution of a two-level system interacting with an environment having the properties of random telegraph signal noise. To give a physical example, this type of a model might be used to describe a spin in presence of a constant intensity magnetic field that changes its sign randomly in time.

One can start to analyze such a system by writing a time-dependent Hamiltonian as
\begin{align}
H(t)=\hbar\sum_{k=1}^3\Gamma_k(t)\sigma_k,
\end{align}
where $\hbar$ is the Planck's constant and $\Gamma_k(t)$ are independent random variables respecting the statistics of a random telegraph signal which can be expressed as $\Gamma_k(t)=a_k n_k(t)$. Here, $n_k(t)$ has a Poisson distribution with a mean equal to $t/2\tau_k$ and $a_k$ is a coin-flip random variable having the values $\pm a_k$. With the help of the von Neumann equation $\dot{\rho}=-(i/\hbar)[H,\rho]$, it is possible to find a solution for the density matrix of the two-level system given by
\begin{align}
\rho(t)=\rho(0)-i \int_0^t\sum_k \Gamma_k(s)[\sigma_k,\rho(s)]ds. \label{isol}
\end{align}
If one substitutes Eq. (\ref{isol}) back into the von Neumann equation and performs a stochastic average, one obtains a master equation of the form
\begin{align}
\dot{\rho}(t)=-\int_0^t\sum_k e^{-(t-t')/\tau_k}a_k^2 [\sigma_k,[\sigma_k,\rho(t')]]dt', \label{sol}
\end{align}
where the memory kernel originates from the correlation functions of the random telegraph signals given as $\langle\Gamma_j(t)\Gamma_k(t')\rangle=a_k^2\exp(-|t-t'|/\tau_k)\delta_{jk}$.

In Ref. \cite{daffer}, Daffer et al. have studied the requirements of completely positive time evolution for this type of a master equation. They have found that the dynamics generated by Eq. (\ref{sol}) is completely positive when two of the $a_k$ are zero, which corresponds to the situation where the colored noise only acts in a single direction. In particular, when the condition $a_3=a$ and $a_1=a_2=0$ is satisfied, the resulting time evolution experienced by the open quantum system is that of a colored noise dephasing channel with non-Markovian features. In this case, the Kraus operators that describes the dynamics of the each two-level system are given by
\begin{align}
M_1 &= \sqrt{[1+\Lambda(\nu)]/2}I_2, \\
M_2 &= \sqrt{[1-\Lambda(\nu)]/2}\sigma_3.
\end{align}
with $\Lambda(\nu)=e^{-\nu}[\cos(\mu\nu)+\sin(\mu\nu)/\mu], \mu=\sqrt{(4a\tau)^2-1}$ and $\nu=t/2\tau$ is the dimensionless time. Since we will consider the dynamics of two-qubits independently interacting with identical colored dephasing environments, we can obtain the dynamical map using Eq. (\ref{Kdynamics}).

We will once again study the problem for initial Bell diagonal states of the form given in Eq. (\ref{Bdiag}). As a result of the dynamics considered here, the Bell diagonal states preserve their general form, and the real three parameters defining them evolve in time as follows
\begin{align}
c_1(\nu)=c_1(0)\Lambda(\nu)^2, \qquad c_2(\nu)=c_2(0)\Lambda(\nu)^2, \qquad c_3(\nu)=c_3(0),
\end{align}
Similarly to the case of Markovian dephasing noise discussed in the previous subsection, one can identify three regions of distinct dynamical features depending on the relations among the initial state parameters $c_1$, $c_2$ and $c_3$ \cite{m3}.

\begin{figure}
\centering
\includegraphics[width=0.65\textwidth]{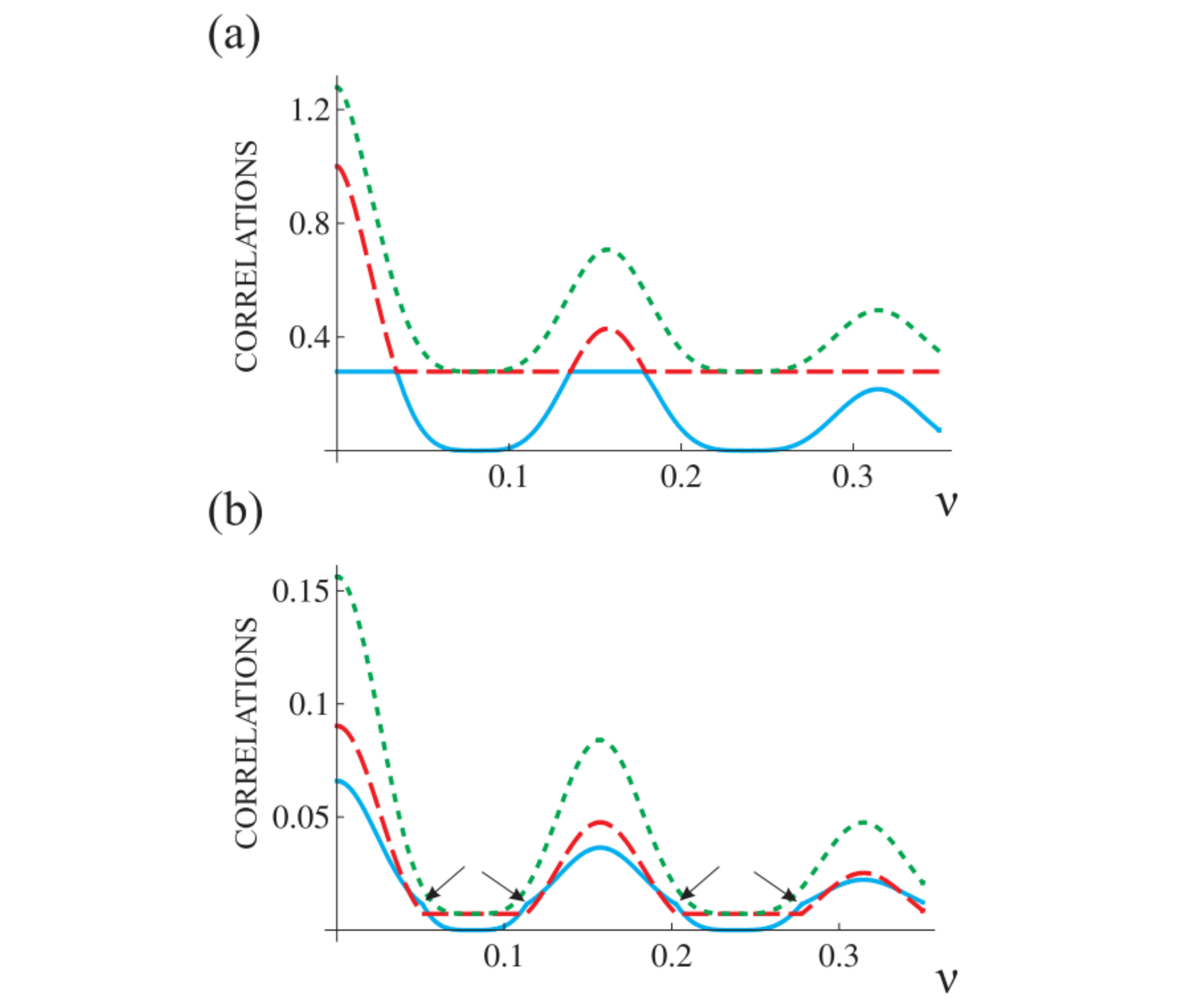} 
\caption{Dynamics of the quantum mutual information (green dotted line), classical correlations
(red dashed line) and quantum discord (blue solid line) as a function of dimensionless time $\nu = t/2\tau$ with $\tau = 5$ and $a=1$. (a) Frozen discord with sudden multiple transitions where initial state parameters are $c_1(0) = 1, c_2(0) = −0.6$ and $c_3 = 0.6.$ (b) Sudden change without frozen behavior with parameters  $c_1(0) = 0.35, c_2(0) = −0.3$ and $c_3 = 0.1$. The arrows show the sudden change points.}
\label{fig4}
\end{figure}

In the first region of parameters where $|c_3(0)|\geq|c_1(0)|,|c_2(0)|$, both the quantum mutual information and the quantum discord display damped oscillations, eventually vanishing asymptotically. There exist no sudden changes in either quantum or classical correlations throughout the time evolution of the open system. On the other hand, classical correlations do not feel the effect of the noise and remain constant at all times, which can be easily seen from Eq. (\ref{clas}) as  $\chi(t)$ is always equal to $|c_3(0)|$ in the course of the dynamics. The second regime includes the time evolution of the initial Bell diagonal states with $c_3=0$. Here, classical, quantum and also total correlations asymptotically tend to zero which means that final state of the open system is a product state, possessing no correlations of either classical or quantum nature. The last and the most interesting region is characterized by the initial states having $|c_3(0)|<|c_1(0)|$ and/or $|c_2(0)|$. Since $\chi(t)$ is initially equal to $\text{max}\{|c_1(t)|,|c_2(t)|\}$, classical correlations begin to decay at first until a certain critical time. At this time point, they become suddenly frozen due to the fact that $|c_3(0)|$, which is constant during the dynamics, becomes greater than $\text{max}\{|c_1(t)|,|c_2(t)|\}$. Recalling in the non-Markovian case that, $c_1(t)$ and $c_2(t)$ are oscillating functions of time, $|c_1(t)|$ or $|c_2(t)|$ might grow bigger than $|c_3(0)|$ one again during the dynamics, in which case classical correlations would start oscillating until they become abruptly constant. In fact, in this last region, we can distinguish two different dynamical behaviours for quantum discord, namely, sudden change dynamics with or without frozen behavior. An example of the first case is shown in Fig. \ref{fig4} (a), where one can observe that quantum and classical correlations display multiple abrupt changes in their decay rates due to the non-Markovian memory effects in the open system dynamics. Indeed, such a behavior can be seen for the whole subclass of Bell diagonal states with $c_{1(2)}(0)=k$, $c_{2(1)}(0)=-c_3 k$ and $c_3(0)=c_3$ with $k$ real and $|k|>|c_3|$. On the other hand, an example of the sudden change without frozen discord behavior is shown in Fig. \ref{fig4} (b). Here, both classical correlations and quantum discord decay until a sudden change point is reached, where classical correlations become constant while quantum discord abruptly changes its decay rate without getting frozen. As time passes, this behavior is repeated with intervals of constant classical correlations.
\vspace{-100cm}
\vspace{100cm}
In order the understand the underlying reason behind the multiple periods of constant discord and sudden changes, one can conduct a geometrical analysis in terms of relative entropy based discord, similarly to what has been done in the previous subsection. Recall that original quantum discord and relative entropy based geometric discord turn out to be equivalent for the Bell diagonal states. Then, in this case, it is possible to track the trajectory of the state under investigation and its closest classical state during the dynamics to understand the geometrical origins of sudden changes and freezing behavior.
\begin{figure}
\centering
\includegraphics[width=0.53\textwidth]{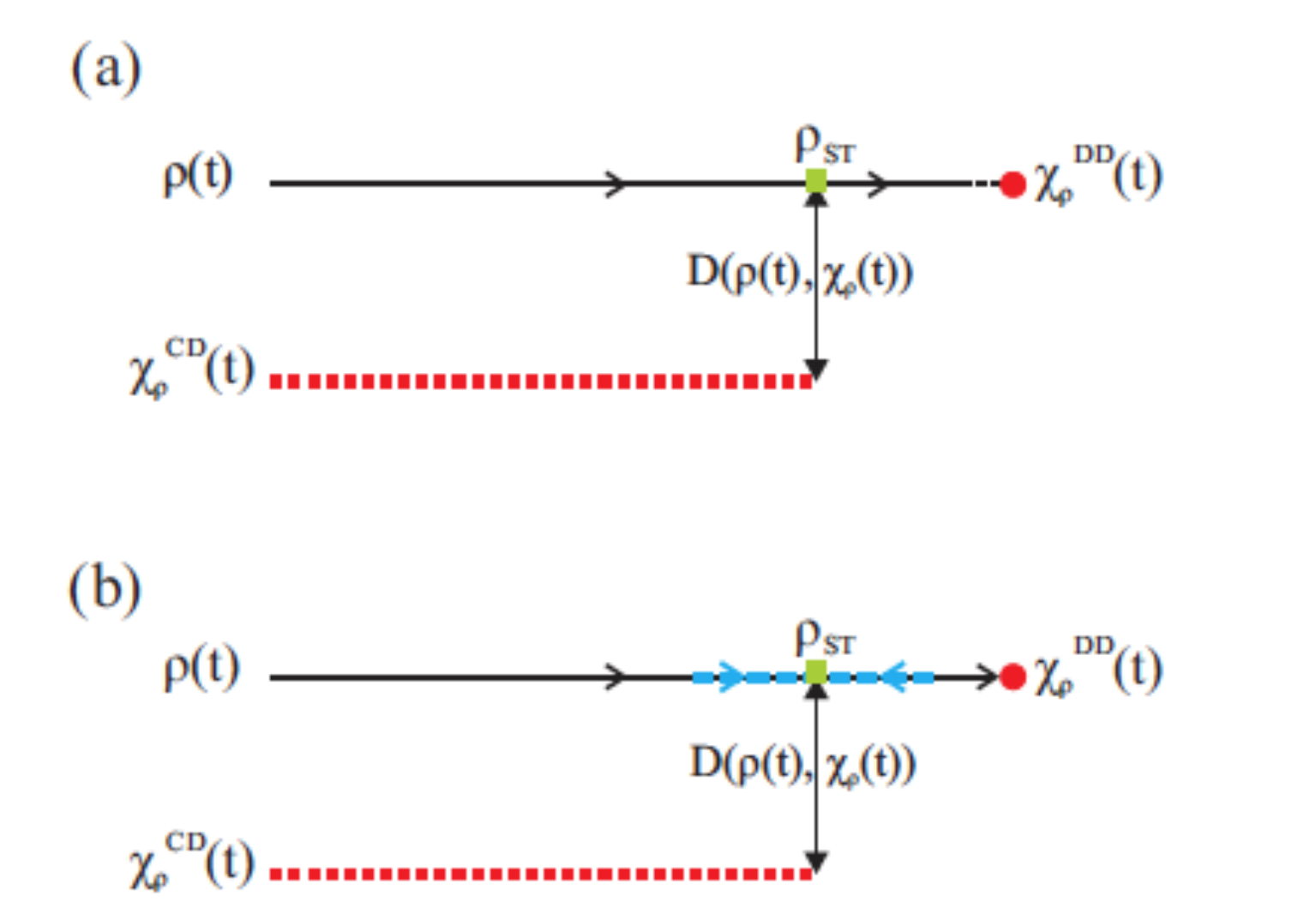} 
\caption{Sketch of the trajectories of the state of the system under investigation and its closest classical state in the Hilbert space in the (a) Markovian, (b) non-Markovian case.}
\label{fig5}
\end{figure}  

Let us first have a look at the Markovian dephasing case illustrated in Fig. \ref{fig5} (a), where there are no oscillations in the dynamics of correlations. Whereas the solid black line represents the trajectory of the considered state $\rho(t)$, the dotted red line depicts the path traced by the closest classical state $\chi_\rho^{CD}(t)$. As can be seen from the figure, the trajectory of the $\rho(t)$ is parallel to the one followed by $\rho^{CD}(t)$, that is, the system has frozen discord until the sudden change point is reached. The green square shows the sudden change point, where the state of the system $\rho_{ST}$ has two closest classical states in equal distance, namely $\chi_\rho^{CD}(t)$ and $\chi_\rho^{DD}(t)$ in correspondence with the eigenvalue crossing point $\lambda_2=\lambda_3$. As the open quantum system keeps evolving in time, it continues to travel along the black line. Therefore, the state under investigation $\rho(t)$ gets closer to its closest classical state $\chi_\rho^{DD}(t)$ asymptotically resulting in decaying quantum discord.

The consequences of the non-Markovian memory effects for the dynamics of the open system can be seen in Fig. \ref{fig5} (b). Note that here the meaning of symbols and trajectories, and the structure of the set of closest classical states remain the same as in Fig. \ref{fig5} (a). The crucial difference is that the state of the system under investigation oscillates around the sudden transition point throughout the dynamics, due to the non-Markovian features of the noise channel. When the considered state $\rho(t)$ first passes through the transition point as it travels from left to right, it reaches the closest classical state $\chi_{\rho}^{DD}(t)$ in a finite time interval. After this point, non-Markovian memory effects force the state $\rho(t)$ to travel back along part of its previous trajectory, which is shown in Fig. \ref{fig5} (b) by the thick dashed blue line partially overlapping with the black one. As it travels back from right to left on the blue line, it once again enters the frozen discord region after crossing the sudden change point. Finally, the direction of path of the state $\rho(t)$ flips once more and it enters the decaying quantum discord region to remain there. Xu et al. observed the dynamics of the correlations in a non-Markovian dephasing environment with an all-optical setup \cite{guo} and found the same features as described here.

\subsection{Time-invariant Discord}
\subsubsection{Time-invariant Discord and Non-Markovian Open Quantum Systems}
The simple but exact characterisation of local pure dephasing non-Markovian noise in qubit systems has lead to a plethora of studies surrounding the quantification and usefulness of non-Markovianity \cite{puremodel1}-\cite{pureeg6}. In this subsection, we examine this exact model in order to reveal the origin of time invariant discord. In more detail, the sudden transition between classical and quantum decoherence, inevitable in Markovian systems, does not necessarily occur in non-Markovian systems. Without this transition, quantum correlations may persist at all times while other dynamical quantities evolve. 

We consider the following microscopic Hamiltonian describing the local interaction of a qubit and a bosonic reservoir in units of $\hbar$ \cite{puremodel1}-\cite{puremodel3}, 
\eq
H=\omega_0\sigma_z+\sum_k \omega_ka^\dagger_ka_k+\sum_k\sigma_z(g_ka_k+g^*a_k^\dagger),
\eeq
with $\omega_0$ the qubit frequency, $\omega_k$ the frequencies of the reservoir modes, $a_k(a_k^\dagger)$ the annihilation (creation) operators and $g_k$ the coupling constant between each reservoir mode and the qubit. In the continuum limit $\sum_k |g_k|^2 \rightarrow\int d\omega J(\omega)\delta(\omega_k-\omega)$ where $J(\omega)$ is the reservoir spectral density. The time local master equation for the qubit is given by,
\eq
\dot\rho=\gamma(t)[\sigma_z\rho\sigma_z-\rho]/2.
\eeq
If the environment is initially in a thermal state at $T$ tempreature, the time-dependent dephasing rate takes the form, 
\eq
\gamma(t)=\int d\omega J(\omega)\coth[\hbar\omega/2k_B T]\sin(\omega t)/\omega
\eeq
resulting in the decay of the density matrix off-diagonal elements, $\rho_{ij}(t)=e^{-\Gamma(t)}\rho_{ij}(0)$, $i\neq j$, with dephasing factor $\Gamma(t)=\int_0^t \gamma(t') dt'$ given by 
\eqa
\Gamma(t)&=&\int_0^\infty d\omega J(\omega) \coth[\hbar\omega/2k_B T][1-\cos(\omega t)]/\omega^2\nn\\&=&\int_0^\infty \: d\omega g(\omega,T)[1-\cos(\omega t)],
\label{Gam}
\eeqa
where the Ohmic spectral densities are given by
\eq
J(\omega)=\frac{\omega^s}{\omega_c^{s-1}}e^{-\omega/\omega_c}
\eeq
with $\omega_c$ the reservoir cutoff frequency. By changing the $s$ parameter, one goes from sub-Ohmic reservoirs ($s<1$) to Ohmic ($s=1$) and super-Ohmic ($s>1$) reservoirs, respectively. Such engineering of the Ohmicity of the spectrum is possible, e.g.,  when simulating the dephasing model in trapped ultracold atoms, as described in Ref. \cite{pureeg2}. A closed analytical expression for the time-dependent dephasing rate can be found in both the zero $T$ and high $T$ limit. In the former case, one obtains, 
\eq
\gamma_0(t,s)=[1+(\omega_c t)^2]^{-s/2}\Gamma_E[s]\sin[s\arctan(\omega_c t)],
\label{gam}
\eeq
with $\Gamma_E[x]$ the Euler gamma function. For high $T$ instead we have, 
\eq
\gamma_\text{HT}(t,s)=2 k_B T\gamma_0(t,s-1)/\omega_c.
\label{HT}
\eeq

Mathematically, the effect of the qubit on its environment is described by a displacement operator acting on each environment mode, with the associated phase conditional on the state of the qubit. The two qubit states excite each mode with opposing phases, leading to a decrease in the overlap between the states of the mode in each case. Destructive interference between excitations of a mode at different times reverses decoherence leading to recoherence at the frequency of the mode; it is the balance between these two effects for different modes, captured by Eq. (\ref{Gam}) that determines whether the dynamics is non-Markovian. From Eq. (\ref{Gam}), a simple link between the onset of non-Markovianity and the form of the reservoir spectrum can be established. As the cosine transform of a convex function is monotonically increasing, a sufficient condition for Markovianity is that $g(\omega,T)$  is convex or, equivalently, the non-convexity of $g(\omega,T)$ is a necessary condition for non-Markovianity. Physically, a convex $g(\omega,T)$ means that any recoherence is always outweighed by more decoherence from lower frequency modes. 

\begin{figure}[t]
\centering
\includegraphics[width=0.4\textwidth]{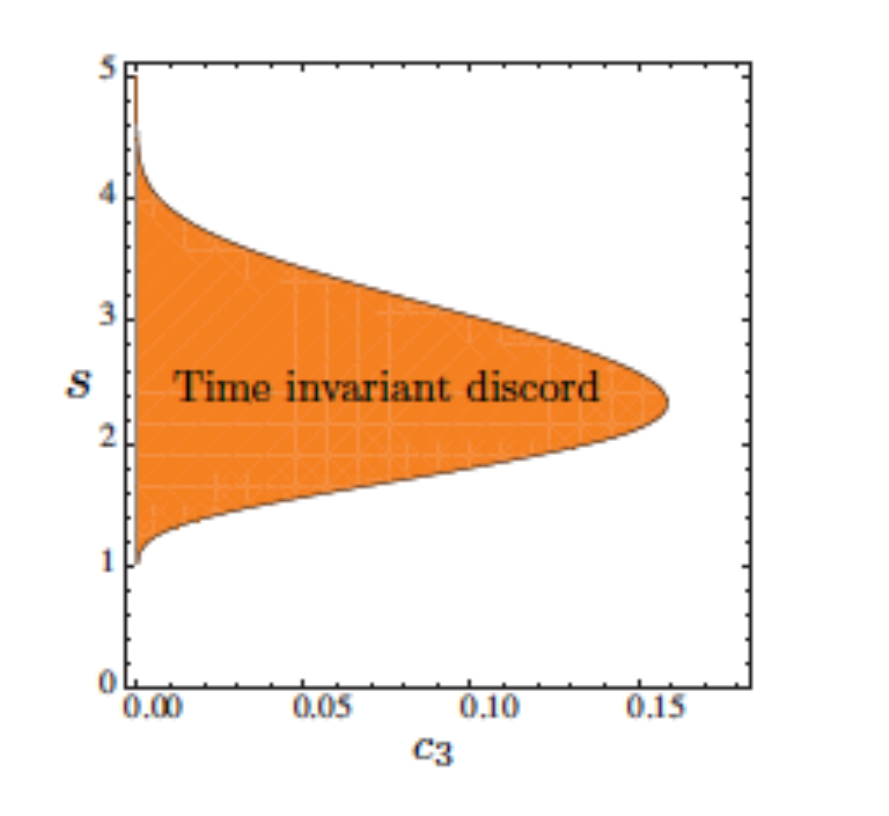}
\caption{The shaded region marks the range of parameters $s$ and $c$ for which the discord is frozen forever
for $T=0$. Outside this region one will always observe a transition from classical to quantum decoherence.}
\label{Pinja1}
\end{figure}

For Ohmic class spectra, a simple connection between the general form of the spectrum and memory effects in the reduced system can be established from the form of the decay rate. Memory effects originate from non-divisible maps, corresponding to dissipators with decay rates which take temporarily negative values. It is straightforward to realise from Eq. (\ref{gam}) that for zero temperature, the dephasing rate takes negative values if and only if $s_\text{crit}>2$. Equally, from Eq. (\ref{HT}), the onset of non-Markovianity for high temperatures is $s_\text{crit}>3$. Hence, for the pure dephasing model, non-Markovianity only occurs for super-Ohmic environments. For intermediate temperatures, $s_\text{crit}$ increases monotonically until $s_\text{crit}=3$ at infinite temperatures. Moreover, it is can be shown that for $s_\text{crit}$ and for all temperatures, the function $g(\omega,T)$ changes from a convex to non-convex function of $\omega$, implying that the condition on the non-convexity of the spectrum is necessary and sufficient for non-Markovianity at all $T$. 

For a system of two qubits in Bell-diagonal states experiencing local dephasing, the dynamics of the mutual information $\mathcal{I}(\rho_{AB}(t))$, the classical correlations $\mathcal{C}(\rho_{AB}(t))$ and the quantum correlations measured by quantum discord $\mathcal{Q}(\rho_{AB}(t))$ are given by Eqs. (14)-(16) with $c_1(t)=c_2(t)=\text{exp}(-2\Gamma(t))$, $c_3(t)=c>0$ and $\chi(t)=\max\{e^{-2\Gamma(t)},c\}$. Hence, one immediately sees that when $e^{-2\Gamma(t)}>c$, classical correlations decay while discord remains constant. On the other hand, if a finite transition time $\tilde t$ such that, 
\eq
e^{-2\Gamma(\tilde t)}=c
\label{cond}
\eeq
exists, then for $t>\tilde t$ the discord starts decaying and the classical correlations remain constant. Contrary to the Markovian dephasing model, the transition time $\tilde t$ now crucially depends not only on initial state of the two-qubit system through the parameter $c$ but also on parameters describing the structure of the reservoir spectral density,  specifically $s$ and the reservoir temperature $T$, through $\Gamma(t)$. 
\begin{figure}[t]
\centering
\includegraphics[width=0.55\textwidth]{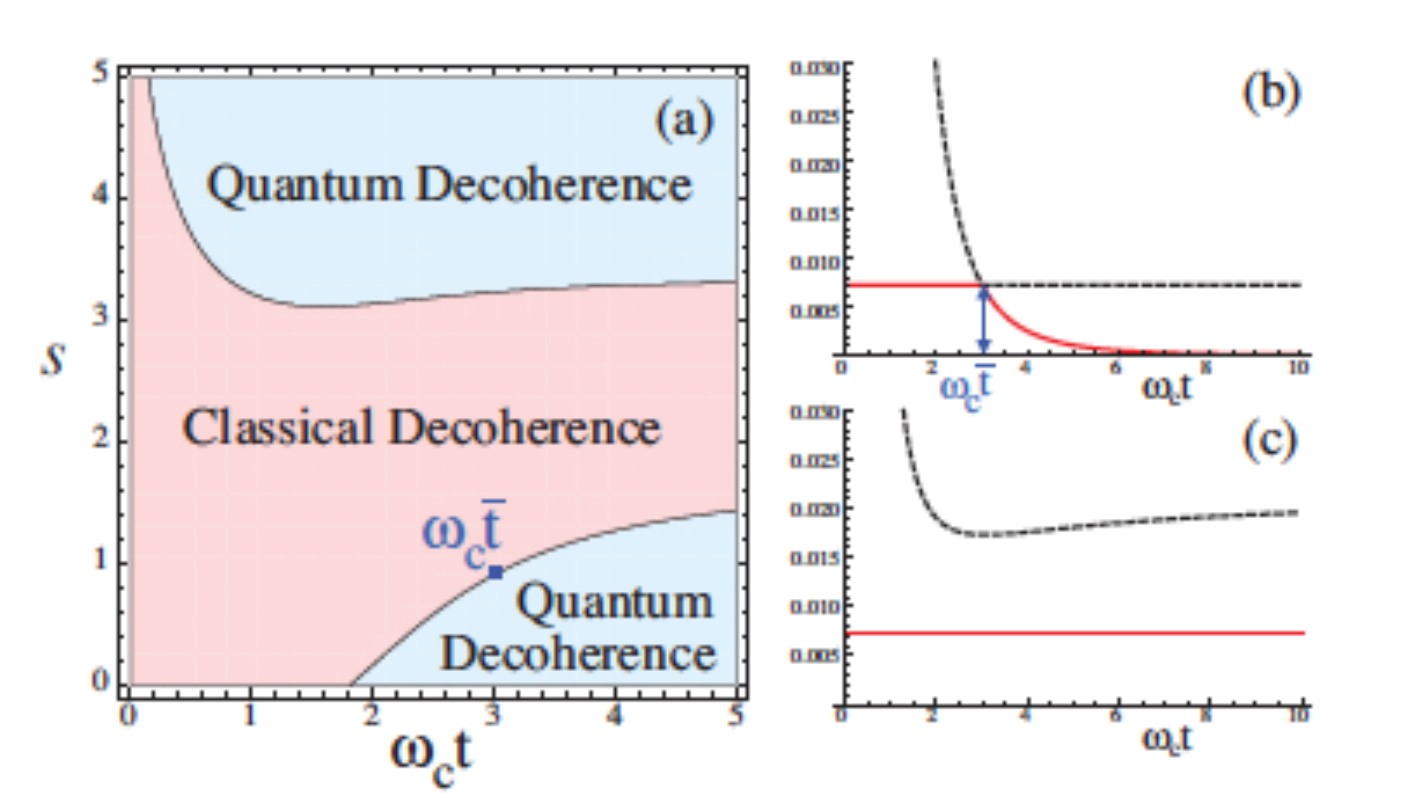}
\caption{(a) Landscape of correlation dynamics in the $s-t$ plane, for $c=0.1$ and $T=0$. Blue areas denote parameters $(t,s)$ corresponding to classical decoherence, red areas to quantum decoherence and the intersection between the two, marking the values of $s$ and $\tilde t$ satisfying Eq. (EQ), defines the transition time $\tilde t$ as a function of the reservoir spectrum parameter $s$. The two insets show discord (solid red line) and classical correlations (dashed black line) for two specific choices of $s$; (b) for $s=1$ the system has a sudden transition from quantum to classical decoherence while for (c) $s=2.5$ the discord is frozen forever. The blue dot in (a) and (b) points the transition time $\tilde t$ for $s=1$.}
\label{Pinja2}
\end{figure}

In Fig. \ref{Pinja1}, the values of $s$ and $c$ for which the condition in Eq. (\ref{cond}) is satisfied are illustrated, for $c=0.1$ and $T=0$. For a certain range of the parameter $s$, Eq. (\ref{cond}) has a solution and accordingly the system has a sudden transition from classical to quantum decoherence at time $\tilde t$. However, there exist a range of values of $s$ for which Eq. (\ref{cond}) has no solution and no transition time $\tilde t$ exists leading to decaying classical correlations, while discord remains frozen forever. The two different cases are illustrated in Fig. \ref{Pinja2} where classical correlations and discord for the Ohmic case $s=1$ and $s=2.5$ are plotted, respectively.  By looking at the asymptotic long time limit of Eq. (\ref{cond}) we can define the $s$ and $c$ parameter space for which time-invariant discord exists (see Fig. \ref{Pinja1}). The value of frozen discord is, 
\eq
\mathcal{Q}=(1+c)\text{log}_2(1+c)/2+(1-c)\text{log}_2(1-c)/2.
\eeq
Therefore, frozen discord takes significant values only for small values of $c$, approximately corresponding to the Ohmic range $2\leq s \leq 3$ when the dynamics is non-Markovian. Increasing the temperature rapidly destroys the time invariant discord phenomenon and in the high temperature limit values of $s$ and $c$ for which this effect occurs can not be found. 

The occurrence of time invariant discord in two qubit dynamics can be connected to the form of the reservoir spectrum and non-Markovianity in the single qubit dynamics. It is straightforward to show that the time invariant discord phenomenon can occur only for reservoir spectra leading to a bounded value of $\Gamma(t)$. This ensures the existence of values of $c$ such that $e^{-2\Gamma(t)}>c$ for all $t$ implying that Eq. (\ref{cond}) is never satisfied. On the other hand, an asymptotic divergence of $\Gamma(t)$ allows for the existence of a transition time $\tilde t$. Such a divergence, and therefore absence of time invariant discord rests on the divergence of $\omega g(\omega, T)$ when $\omega\rightarrow 0$ occurring for $s\leq 1(2)$ at zero (finite) temperature. Futhermore, convexity and thus Markovianity is ensured if $g(\omega, T)$ diverges at low frequencies, occurring for $s\leq s_\text{crit}=2(3)$. Hence, time invariant discord and non-Markovianity are intimately related and ultimately rely on the eventual dominance of recoherence over decoherence, thus both require the suppression of coupling to low frequency modes, embodied by the low frequency dependences of $J(\omega)$ and $g(\omega, T)$.

\subsubsection{Time-Invariant Discord in Dynamically Decoupled Systems}
We now continue our study of the exact purely dephasing system in order to reveal the connection between time invariant discord and optimal control in the form of dynamical decoupling. In more detail, we now compare reservoir engineering techniques based on a modification of the reservoir spectral density through the Ohmicity parameter in order to change the Markovian character of the dynamics with reservoir engineering exploiting dynamical decoupling (which in turn, can also be seen as effective filtering of the spectral density). Dynamical decoupling (DD) techniques for open quantum systems are considered one of the most successful control protocols to suppress decoherence in qubit systems \cite{ref1Lorenza, ref2Lorenza}. Inspired by the spin-echo effect, dynamical decoupling involves the application of a sequence of external pulses to the system which induce unitary rotations in order to counter the harmful effects of the environment \cite{dd1}--\cite{dd5}. Specifically, ``bang bang" periodic dynamical decoupling (PDD) schemes have been shown to prolong coherence times and restore  decaying correlations in quantum systems which are undergoing decoherence \cite{dd2}. 

We now recall the exact dynamics obtained in Ref. \cite{Paper} which address the purely dephasing qubit behavior in the presence of an arbitrary sequence of instantaneous bang-bang pulses. Each pulse is modelled as an instantaneous $\pi$-rotation around the $x$-axis. An arbitrary storage time is considered, $t$, during which a total number of $N$ pulses are applied at instants $\{t_1,...t_n,...t_f\}$, with $0<t_1<t_2<...<t_f<t$. As shown by Uhrig \cite{U}-\cite{Ua}, the controlled coherence function $\Gamma(t)$ can be worked out as,
\eq
\Gamma(t)= \left\{ 
  \begin{array}{l l}
    \Gamma_0(t) & t\leq t_1 \\
    \Gamma_n(t) & t_n<t\leq t_{n+1},0<n<N \\
    \Gamma_N(t) & t_f<t \label{eq:Gamma} 
  \end{array} \right.,
 \eeq
where the exact representation of the controlled decoherence function $\Gamma_n(t)$ for $1\leq n \leq N$, can be written in the following form:
\eq
\Gamma_n(t)=\int_0^\infty \frac{J(\omega)}{2\omega^2}|y_n(\omega t)|^2d\omega, \:\:\:\:\:n\geq0,
\label{gn}
\eeq
Further, from Eq. (\ref{gn}), $|y_0(\omega t)|^2=|1-e^{i\omega t}|^2$, and 
\eq
y_n(z)=1+(-1)^{n+1}e^{iz}+2\sum_{m=1}^n(-1)^me^{iz\delta_m}, \:\:\:\:\:z>0.
\eeq
Here, it is understood that the $n$th pulse occurs at time $t_n=\delta_nt$ and $0<\delta_1<...<\delta_n<...<\delta_s<1$. In order to express the controlled decoherence function $\Gamma_n(t)$ in terms of its uncontrolled counterpart $\Gamma_0(t)$, we simply relate $|y_1(\omega t)|^2$ to $|y_0(\omega t)|^2$ to write,
\eq
\Gamma_1(t)=-\Gamma_0(t)+2\Gamma_0(t_1)+2\Gamma_0(t-t_1). 
\eeq
Upon iteration, and relating again, $|y_n(\omega t)|^2$ to $|y_0(\omega t)|^2$, we find the decoherence rate for $n$ pulses, 
\eqa
\Gamma_n(t)&=&2\sum_{m=1}^n(-1)^{m+1}\Gamma_0(t_m)\nn\\&+&4\sum_{m=2}^n\sum_{j<m}\Gamma_0(t_m-t_j)(-1)^{m-1+j}\nn\\&+&2\sum_{m=1}^n(-1)^{m+n}\Gamma_0(t-t_m)+(-1)^n\Gamma_0(t).
\label{Gamma}
\eeqa

It is well known that the performance of dynamical decoupling techniques crucially depends on the temporal separation of the pulses. Moreover, the performance can be linked to the timescale of the environment correlation function, highlighting the important role played by spectral properties of the noise causing decoherence and introducing errors \cite{Daniel1}. Following this, in Ref. \cite{Pub3}, an explicit connection between the direction of information flow and dynamical decoupling is established. The influence of dynamical decoupling on the direction of information flow can be seen directly the following relation, connecting $\gamma_n(t_n)$, i.e., $\dot\Gamma_n(t_n)$ at the moment $t_n$ when the system is pulsed and the corresponding quantity at the previous instant: 
\eq
\gamma_n(t_n)= -\gamma_{n-1}(t_n),
\label{gammannm1}
\eeq
where for $1\leq n \leq N$, 
\eq
\gamma_n(t)=2\sum_{m=1}^n (-1)^{m+n}\gamma_0(t-t_m)+(-1)^n\gamma_0(t).
\label{rev}
\eeq
From this expression, it is immediate to associate the change of sign of the decay rate at time $t_n$ to a reversal of information flow occurring at the instant the system is pulsed.  As a direct consequence, a Markovian open system dynamics will always be transformed to a non-Markovian dynamics. Hence, from a reservoir engineering perspective, in order for each pulse to result in revivals in dynamic quantities, one should choose environments with Ohmicity parameters $s<2$, corresponding to unperturbed Markovian dynamics. 

It is well established that the use of either DD techniques or non-Markovian effects prolongs the preservation of both entanglement and discord in the presence of environmental noise \cite{1}-\cite{5}. Unfortunately, as realised from Eq. (\ref{rev}), simultaneous use of non-Markovian reservoir engineering and DD protocols is counterproductive for avoiding decoherence. In more detail, if a pulse occurs during an interval of information back flow, information flow is reversed, inducing rapid decay in dynamic quantities such as coherence. We now consider the preservation of quantum and classical correlations and the possibility of creating time invariant discord through PDD techniques in relation to the initial conditions (unperturbed dynamics). For a more detailed insight, we consider two different scenarios of local dephasing noise in the presence of DD. Specifically, we firstly consider the scenario where both qubits $A$ and $B$ interact locally with identical dephasing environments, where in this case, both qubits are subject to pulsing. Secondly, we consider the case where only one of the qubits interacts with the dephasing environment, while the other qubit is entirely protected from decoherence (the expressions for classical and quantum correlations hold in this case provided we exchange $e^{-\Gamma(t)}$ with $e^{-2\Gamma(t)}$). We note that only the qubit experiencing noise is subject to DD. Without the possibility of analytically or numerically defining continuously pulsed dynamics in the asymptotic time limit, we focus on discord which remains ``time-invariant" within a chosen experimental time interval rather than forever. 

\begin{figure}[t]
\centering
\includegraphics[width=0.6\textwidth]{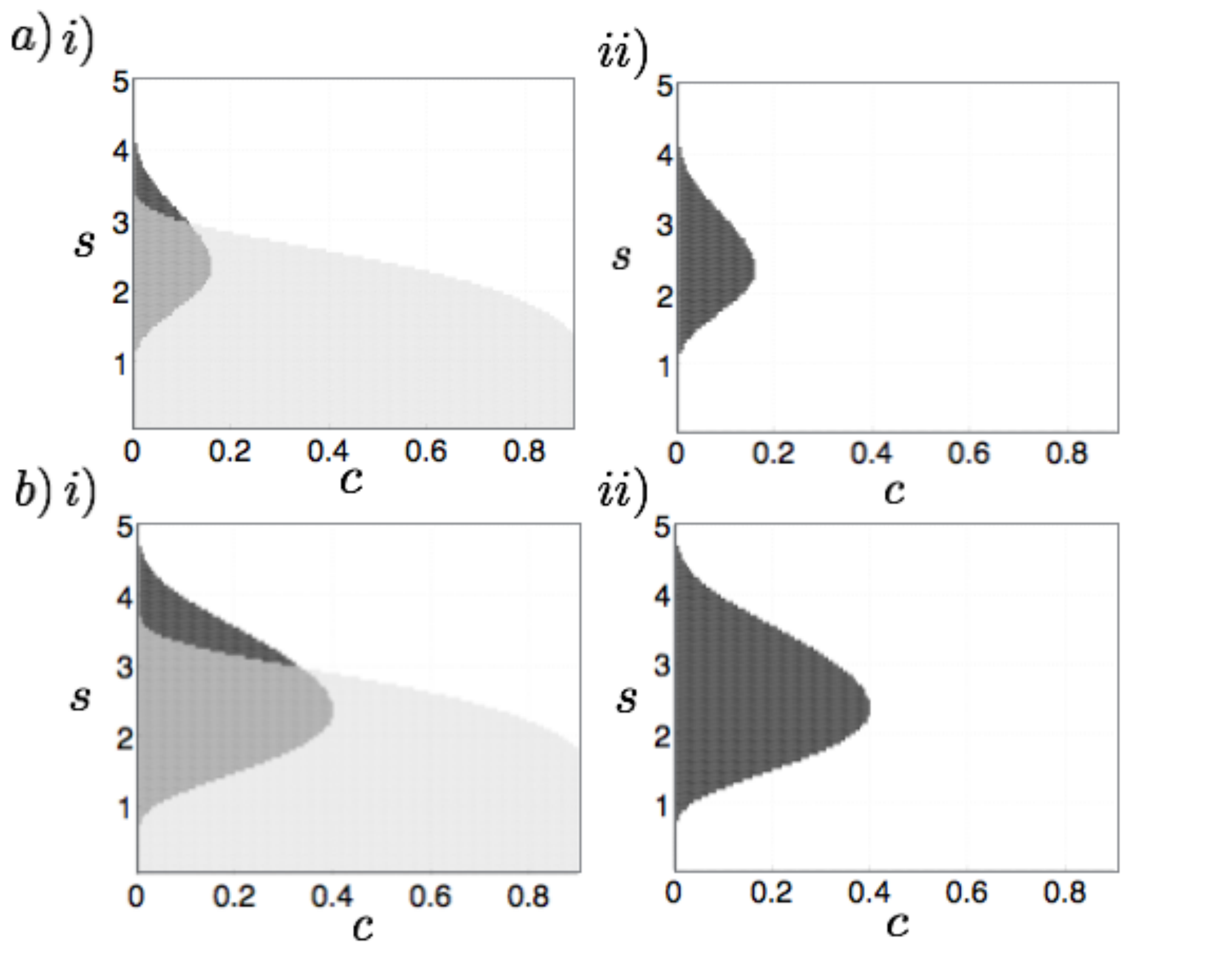}
\caption{The black region shows the range of $s$ and $c$ parameters for which the discord is frozen forever for a free system evolving without pulsing. The grey region shows the range of $s$ and $c$ for dynamical decoupled systems with small interval spacing $\Delta t=0.3\omega_c^{-1}$ (i) and long interval spacing $\Delta t=3\omega_c^{-1}$ (ii).
The plots in (a) are for the case where both qubits are affected by noise, as the plots in (b) are for single qubit noise. Outside these regions, one will always observe a transition from classical to quantum decoherence and thus no time-invariant discord. The final pulse is applied at $t_f=N_\text{max}\Delta t\leq 25\omega_c^{-1}$ where $N_\text{max}$ is the maximum number of pulses that can be applied within the time interval $0\leq t\leq 25\omega_c^{-1}$. Quantities plotted are dimensionless. }
\label{A11}
\end{figure}

We now compare the regions of $s$ and $c$ for which time-invariant discord exists for both the pulsed and unperturbed case. For a short pulse $\Delta t=0.3\omega_c^{-1}$, we see from Fig. \ref{A11} a,b i), for two-sided and one-sided noise respectively, time invariant discord is created for a wider range of parameters compared to the unperturbed case. Specifically, time-invariant discord is created for sub-Ohmic values of $s$, (i.e., $s<1$) for up to very high values of $c$, only when the system is subject to DD. Moreover, increasingly significant values of time invariant discord, corresponding to increasing values of $c$, occur for $s<2$. These conclusions are independent of the specific pulse interval chosen provided that $\Delta t<\bar t$ where $\bar t$ defines the first time instant information back flow occurs for the unperturbed dynamics. For $s\lesssim 1$, coherence is maintained close to unity with no degradation shown to occur within computable times. Hence, one can conjecture that time invariant discord will be created in the asymptotic long time limit ($t_f\rightarrow\infty$). On the other hand, for $s> 1$, as $t_f$ increases, the region of time-invariant discord will decrease as the coherence decays to increasingly small values. Physically, it is self-evident that the regions of invariant discord, in the absence and presence of control, become larger when one of the qubits is fully protected against noise. We point out however that the difference is less pronounced between the one-sided and two-sided noise for discord created with DD pulses and more significant in the case of unpulsed non-Markovian dynamics. 

As the pulse interval $\Delta t$ increases, the overlap between the regions of uncontrolled and controlled time-invariant discord becomes smaller. In Fig. \ref{A11} a ii) b ii), the destructive influence of DD on time-invariant discord for large pulse intervals is immediately evident. Indeed, time-invariant discord is completely destroyed for both one-sided and two-sided noise.  Hence, we find that short pulse interval DD schemes paired with Markovian environments (specifically $s<1$), are optimal for the creation of time-invariant discord for a larger range of initial states when compared to strategies relying on non-Markovianity alone as a resource.  Following this, one can say that relying only on non-Markovianity as a resource of time-invariant discord becomes more preferable in general as the pulse interval $\Delta t$ increases. 

\subsection{Conclusions}
In this chapter, we have presented a comprehensive exploration of the phenomena of frozen and time invariant quantum discord for bipartite quantum systems evolving under several different open quantum system models. We have discussed the suitable conditions for the initial states and the properties of the considered environmental models so that the behaviours of frozen or time invariant discord can be consistently observed. We have also considered non-Markovian open system models and shown how the effects of the memory in the dynamics affect the time intervals during which quantum discord becomes frozen. Finally, we have elaborated on the effect of pulsed dynamical decoupling techniques on the preservation of quantum discord at all times during the dynamics.

\subsection*{Acknowledgments}{S.M. acknowledges the Horizon 2020 EU collaborative project QuProCS (Grant Agreement 641277), the Academy of Finland (Project no. 287750)  and the Magnus Ehrnrooth Foundation. C.A. acknowledges financial support from the EPSRC (UK) via the Doctoral Training Centre in Condensed Matter Physics. G. K. is grateful to Sao Paulo Research Foundation (FAPESP) for the fellowship given under grant number 2012/18558-5.}

\end{document}